\documentclass{elsart}

\usepackage{graphicx}

\def\trademark{{\ooalign{\hfil\raise.07ex\hbox{\scriptsize R}\hfil\crcr\mathhexbox20D}}}
\def\sep{-}
\def\saccar{\textit{Saccharomyces cerevisi{\ae}}}

\journal{Biophysical Chemistry}

\begin{document}

\begin{frontmatter}

\title{Model evaluation for glycolytic oscillations in yeast 
biotransformations of xenobiotics}
\author[INSA,MPI]{L. Brusch\corauthref{cor}\thanksref{thanks1}},
\thanks[thanks1]{LB is supported by the Max Planck Society through
       an Otto Hahn fellowship.}
\corauth[cor]{Corresponding author.}
\ead{brusch@mpipks-dresden.mpg.de}
\author[Reg,MPI]{G. Cuniberti\thanksref{thanks2}},
\thanks[thanks2]{GC research is funded by the Volkswagen Foundation.}
\author[TUD]{M. Bertau}
\address[INSA]{Centre de Bioing{\'e}nierie Gilbert Durand INSA-DGBA, \\
       Av. de Rangueil 135, F-31077 Toulouse Cedex 4, France}
\address[MPI]{Max-Planck-Institut f\"ur Physik komplexer Systeme,  \\
       N\"othnitzer Stra{\ss}e 38, D-01187 Dresden, Germany}
\address[Reg]{Institute for Theoretical Physics, University of Regensburg, \\
       D-93040 Regensburg, Germany}
\address[TUD]{Institut f\"ur Biochemie, Technische Universit\"at Dresden, \\
       D-01062 Dresden, Germany}

\begin{abstract}
Anaerobic glycolysis in yeast perturbed by the
reduction of xenobiotic ketones is studied numerically in
two models which possess the same topology 
but different levels of complexity.
By comparing both models' predictions for concentrations and fluxes as well as 
steady or oscillatory temporal behavior we answer the question what phenomena
require what kind of minimum model abstraction.
While mean concentrations and fluxes are predicted in agreement by both models 
we observe different domains of oscillatory behavior in parameter space.
Generic properties of the glycolytic response to ketones are discussed.
\end{abstract}

\end{frontmatter}

\section{Introduction}

The reductionist attempt to identify the function of all genetic constituents 
of a living organism is one of the major goals of the post\sep genomic 
research~\cite{Oliver96,MR96}.
In a living cell, however, single molecular constituents are linked together in 
a \textit{complex network} including gene regulation, signaling and
metabolic pathways~\cite{Lee02,Michal98}.
Whether the detailed knowledge of the individual properties of metabolites 
taking part in enzymatic reactions may help to infer the large scale behavior 
of the living cell is another question challenging life\sep scientists, and
its answer will necessarily imply a highly inter\sep disciplinary cooperative 
effort.

A recent study of gene regulatory networks for early fruit fly development 
has suggested that the inherited robustness of the network's behavior 
may solely result from the network's topology~\cite{Odell00}. 
This robustness may render the observed behavior insensitive to the quantitative 
details of the microscopic interactions. 
Hence modeling of such networks was successful even for rather crude
approximations of the details.
Other findings on artificial gene circuits showed a broad variety of
different functions despite the same topology, suggesting that quantitative
details become important for some topologies \cite{Leibler02}.
Moreover, the construction of a mathematical model and the corresponding 
abstraction of
the complex intracellular network are always confronted with the problem 
of neglecting ``unnecessary'' details. 
The issue of the appropriate level of abstraction has been disputed particularly
in the case of metabolic networks.
Therein it has been suggested that modeling efforts risk to be no longer taken 
serious by experimentalists if modeling is continued by means of ``too'' strong 
abstractions or falsified toy models~\cite{Faraday01}.

\begin{figure}[ht]
\centerline{\includegraphics[width=.6\textwidth]{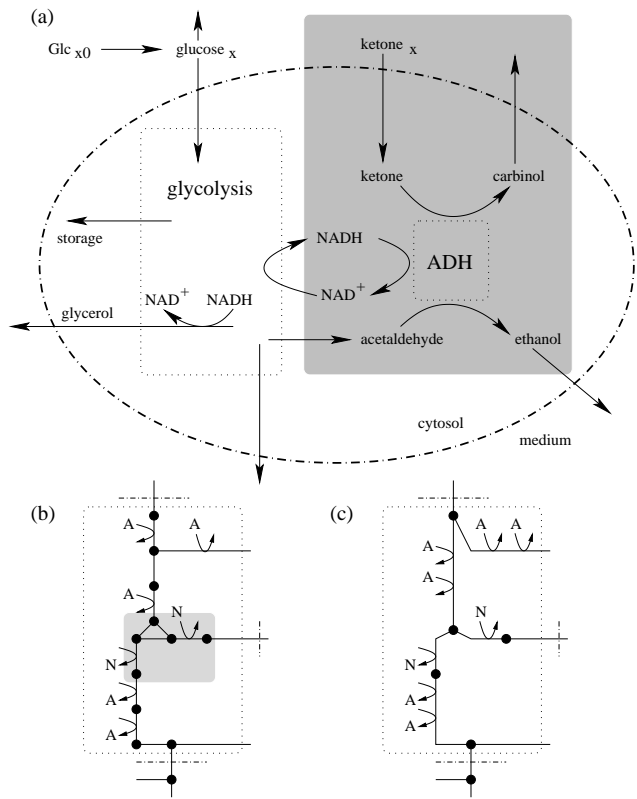}}
\caption{\label{fig:sketch} 
Schematic view of a yeast cell in chemostat culture and models of different
degree of abstraction. 
(a) Glucose influx at constant flow rate is parameterized by the concentration 
[Glc$_x$]$_0$. 
The perturbation (dark shaded box) is controlled by the concentration [K$_x$]
of ketone in the medium and described by the same set of equations 
with identical parameter values for two models.
The full scale model (b) and core model (c) describe glycolysis (dotted box)
at different degrees of complexity.
Dots (lines) denote metabolites (reactions), ``N'' (``A'') stands for 
NADH (ATP) and the light shaded box in (b) is considered in detail in 
Subsec.~\ref{subsec:phaseshift}.
}
\end{figure}

Here we compare predictions from two models of different levels of abstraction 
to characterize the consequences of neglecting details.
For this fundamental study we choose the metabolic network of glycolysis 
(Fig.~\ref{fig:sketch})
in baker's yeast (\saccar) which is of particular interest to 
biotechnology~\cite{OON00,S96}
and models of different degree of complexity have been suggested and used
\cite{Rizzi97,DSH99,WH00,WPSSHW00,TPREvdWSWBvDWS00,Hynne01,RWKS02}.
We choose two models that translate the metabolic network
into systems of ordinary differential equations (ODEs),
a procedure that is well established \cite{HS96,TCN01,FMWT02}.
Our subsequent analysis treats the ODEs fully nonlinear and
focuses on the temporally asymptotic, self\sep organising behavior of the models.
The distinction between steady and oscillatory states in the models appears to 
be crucial, since several physiological advantages of the latter have been 
proposed.
The dependences of the concentrations and fluxes on external control
parameters are numerically computed via a bifurcation analysis \cite{S94,KG95}.
This approach enables detailed studies of large parameter spaces that would be
more time consuming by simulation of the complete time courses including 
transient dynamics \cite{Mendes97}.
This issue will gain importance as the analysed theoretical models increase in
size and complexity.

In order to test the predictive power of both models,
we perturb both models by the same set of simple equations that account for
the conversion of a xenobiotic (here we choose a ketone) by a single enzyme
(here alcohol dehydrogenase) into a chiral carbinol
which consumes the cofactor nicotinamide adenine dinucleotide (NADH), 
see Fig.~\ref{fig:sketch}(a). 
This particular perturbation was chosen both for its pharmaceutical relevance
and its broad impact on the backbone of glycolysis via the NADH mediated 
interference with the redox balance of the cell. 
The pharmaceutical interest stems from the increasing demand for 
stereoisomerically pure pharmacologically active compounds (PACs).
The stereo-geometry of PACs essentially determines the physiological effects of a 
chiral drug, as the thalidomide (contergan\textsuperscript{\small\sf\trademark}) 
tragic showed in the 1950s~\cite{CF00}.
Stereoisomerically pure carbinols are central precursors of modern, 
innovative pharmaceuticals and make for more than 
50\% of all essential substructures in pharmaceuticals~\cite{S96}.
Today stereo-selective carbinol synthesis by means of biotransformation, 
{\it i.e.} chemical reactions catalyzed by enzymes {\it in vivo}, 
is indispensable for the production of PACs \cite{BBWH01,DFMPT99,CBLA95}.
Further progress in the understanding and rational manipulation of 
biotransformatory carbinol syntheses
strongly depends on mathematical models that account for the 
limiting reproduction of the cofactor (NADH in the considered case) and the
underlying regulatory networks~\cite{B01,B02}.
The models studied in this paper also constitute a first step in this direction.
In the future we will extend the analysis to other eucaryote's metabolic 
responses to a range of pharmacologically active compounds.

We emphasize the exposure of systems in oscillatory states to xenobiotics as a 
sensitive experimental probe of the network's global response. 
The global response is experimentally accessible via 
NADH\sep fluorescence, and oscillatory or steady behavior is easily
distinguishable as individual cells synchronize their oscillations~\cite{DSH99}.
The temporal frequency of the oscillations can be 
measured with high accuracy.

The paper is organized as follows. 
In Sec.~\ref{theory} two models of glycolysis and the unique set of equations
representing the perturbation are introduced and the employed numerical method 
of bifurcation analysis is outlined.
Section~\ref{results} presents our results which are discussed in 
Sec.~\ref{conclusions}.


\section{Models and Methods}\label{theory}

The glycolytic pathway of \saccar\ fermenting glucose to ethanol is one of the 
best studied metabolisms from both the experimental and modeling 
side~\cite{Rizzi97,DSH99,WH00,WPSSHW00,TPREvdWSWBvDWS00,Hynne01,RWKS02}.
We consider two different mathematical models in terms of ODEs that identically 
represent the topology of the glycolytic pathway under anaerobic conditions, 
see Fig.~\ref{fig:sketch}(b),(c).

\subsection{Glycolysis}

Specifically we start from the full scale model of glycolysis devised by 
Hynne {\it et al.} \cite{Hynne01} 
and compare its predictions with those of a core model with
the same topology adapted from Wolf {\it et al.} \cite{WH00}.
Public domain ``Silicon Cell'' versions of the models can be found at
{\tt http://www.jjj.bio.vu.nl}, including a graphical overview and the
mathematical expressions.
The full scale model contains 22 variables for concentrations of involved metabolites
and quantitatively accounts for most known details on enzyme regulation
in order to precisely describe the supercritical onset of oscillations as 
observed experimentally~\cite{DSH99,Hynne01}.
The core model has been derived from the original 9
variable model \cite{WH00} by addition of the variable
$S_1^{ex}$ (extracellular glucose) and the corresponding 
flux balance of the medium 
$j_0^{ex} = y_{\mathrm{vol}} k ([{\mathrm{Glc}}_x]_0-S_1^{ex})$ 
and a simple form of glucose transporter saturation 
$j_0 = k_0 (S_1^{ex}-S_1)/(S_1^{ex}-S_1+K_{\mathrm{trans}})$
as well as lactonitrile formation 
$j_{\mathrm{7CN}} = y_{\mathrm{vol}} k_{\mathrm{CN}} S_4^{ex}$ 
in the presence of cyanide.
A storage flux $j_{\mathrm{Store}} = k_{\mathrm{Store}} S_1 A_3$ is included
in analogy with the full scale model but
as a consequence of combining the three reactions downstream of glucose,
$j_{\mathrm{Store}}$ is fed by intracellular glucose $S_1$ in the core model
instead of glucose-6-phosphate in the full scale model.
The consumption of two subsequent entities ATP per glucose for storage would
contribute with $-2 j_{\mathrm{Store}} =-2 k_{\mathrm{Store}} S_1 A_3$ to the 
ATP balance. 
However, we find that this term overestimates the feedback of a varying glucose
concentration on the ATP balance and we always observed subcritical onsets of 
oscillations under this assumption. 
Indeed, not intracellular glucose but glucose-6-phosphate is the substrate
and allosteric activator of the pathway to carbohydrate storage \cite{JMF01}.
In the ATP balance, we therefore keep the ATP consumption for storage included
in the term $-k_5 A_3$ of unspecific ATP consumption in line with the original
9 variable model \cite{WH00}.
The remaining nomenclature and kinetic terms are identical with 
Ref.~\cite{WH00} but 
we have fitted a new set of parameters (see Table \ref{tab:par}) 
to describe the same data as the full scale model.

\begin{table}
\begin{center}
\begin{tabular}{lrr}
\hline
\scriptsize parameter & \scriptsize value in present model & \scriptsize value in model of Ref.~\cite{WH00} \\
\hline  
\scriptsize $y_{\mathrm{vol}}\cdot k$ (min$^{-1}$)& \scriptsize 59 $\cdot$ 0.048& \scriptsize 1.3 \\[-3mm]
\scriptsize $k_1$ (mM$^{-1}$ min$^{-1}$)& \scriptsize 6211.86& \scriptsize 100.0 \\[-3mm]
\scriptsize $k_2$ (mM$^{-1}$ min$^{-1}$)& \scriptsize 38.3756& \scriptsize 6.0 \\[-3mm]
\scriptsize $k_3$ (mM$^{-1}$ min$^{-1}$)& \scriptsize 126819& \scriptsize 16.0 \\[-3mm]
\scriptsize $k_5$ (min$^{-1}$)& \scriptsize 22.2857& \scriptsize 1.28 \\[-3mm]
\scriptsize $k_6$ (mM$^{-1}$ min$^{-1}$)& \scriptsize 6.93908& \scriptsize 12.0 \\[-3mm]
\scriptsize $\kappa$ (min$^{-1}$)& \scriptsize 24.7& \scriptsize 13.0 \\[-3mm]
\scriptsize $q$  &  \scriptsize 4.0&   \scriptsize 4.0 \\[-3mm]
\scriptsize $K_I$ (mM) & \scriptsize 0.52&   \scriptsize 0.52\\[-3mm]
\scriptsize $N_{\mathrm{total}}$  (mM) & \scriptsize 0.98& \scriptsize 1.0 \\[-3mm]
\scriptsize $A_{\mathrm{total}}$  (mM) & \scriptsize 3.6 & \scriptsize 4.0 \\[-3mm]
\scriptsize $k_0$ (mM min$^{-1}$)& \scriptsize 48.1& \scriptsize - \\[-3mm]
\scriptsize $K_{\mathrm{trans}}$ (mM) & \scriptsize 0.002 & \scriptsize - \\[-3mm]
\scriptsize $k_{\mathrm{Store}}$ (mM$^{-1}$ min$^{-1}$) & \scriptsize 17.1& \scriptsize - \\[-3mm]
\scriptsize $k_{\mathrm{CN}}$   (min$^{-1}$) & \scriptsize 0.01477& \scriptsize - \\[-3mm]
\scriptsize $ADH_{\mathrm{total}}$ (mM) & \scriptsize 0.01& \scriptsize - \\[-3mm]
\scriptsize $k^{\mathrm{ADH}}_1$ (mM$^{-2}$ min$^{-1}$) & \scriptsize 80800&  \scriptsize - \\[-3mm]
\scriptsize $k^{\mathrm{ADH}}_2$ (min$^{-1}$) & \scriptsize 1000&   \scriptsize - \\[-3mm]
\scriptsize $k^{\mathrm{ADH}}_3$ (min$^{-1}$) & \scriptsize 5400&   \scriptsize - \\
\hline
\end{tabular}
\end{center}
\caption{Comparison of parameters for our core model with the original model 
(Reference) by Wolf {\it et al.} \cite{WH00}. The values of $y_{\mathrm{vol}}, 
k, \kappa, N_{\mathrm{total}}, A_{\mathrm{total}}, ADH_{\mathrm{total}},
k^{\mathrm{ADH}}_{1\ldots 3}$ are identical in 
both models and taken
(except for $ADH_{\mathrm{total}}$ and $k^{\mathrm{ADH}}_{1\ldots 3}$)
from Ref.~\cite{Hynne01}.
For the full scale model, parameters are
identical with Hynne {\it et al.} \cite{Hynne01}.}
\label{tab:par}
\end{table}

\begin{figure}
\vspace*{2mm}
\centerline{\includegraphics[width=.99\textwidth]{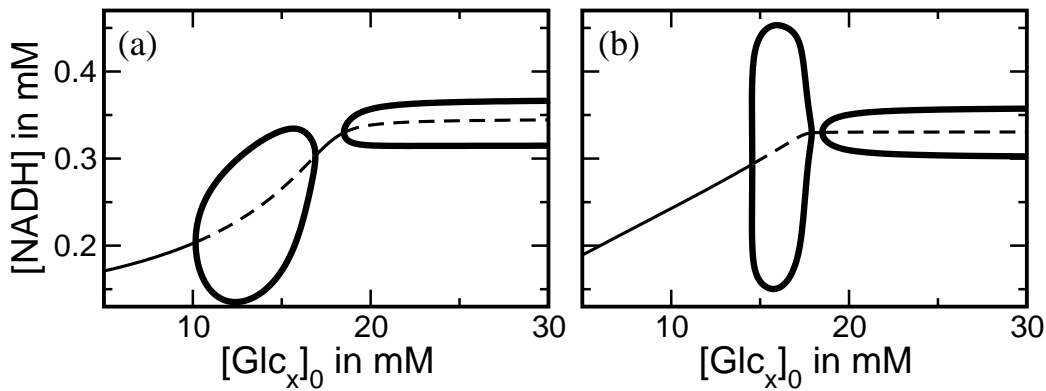}}
\caption{\label{fig:bifglc} 
Bifurcation diagrams of (a) the full scale model and (b) the core model
of glycolysis showing steady states (thin curves) as well as minimum and maximum
(thick) of NADH oscillations for varied inflow glucose concentration
[Glc$_x$]$_0$. Solid (dashed) curves represent stable (unstable) states.
}
\end{figure}

In both models, we have replaced the reduction of acetaldehyde (ACA) 
to ethanol (EtOH) by the same simple reaction scheme:
\begin{equation}
\begin{array}{lc}
\mbox{ACA + NADH + ADH} &
\begin{array}{c}
k^{\mathrm{ADH}}_1 \\[-3mm]
\rightleftharpoons \\[-2mm]
k^{\mathrm{ADH}}_2
\end{array}\\[5mm]
\mbox{ADH$_{\mathrm{ACA}}$} \qquad 
\begin{array}[b]{c}
k^{\mathrm{ADH}}_3 \\[-4mm]
\to
\end{array} \qquad 
\mbox{ADH + NAD$^+$ + EtOH}
\end{array}
\label{equ:adh_aca}
\end{equation}
which later enables the coupling with a second reaction that utilizes the same 
enzyme alcohol dehydrogenase (ADH; EC 1.1.1.1).
We introduce the concentration [ADH$_{\mathrm{ACA}}$] of the bound enzyme 
as an additional variable with the ODE
\begin{equation}
\frac{\rm d [{\mathrm{ADH}}_{\mathrm{ACA}}]}{\rm dt} = v_{\mathrm{ADH,ACA}}^+ - v_{\mathrm{ADH,ACA}}^- ~.
\label{equ:adh_aca_ode}
\end{equation}
Its gain ($v_{\mathrm{ADH,ACA}}^+$) and loss ($v_{\mathrm{ADH,ACA}}^-$) fluxes 
are described by 
simple mass action kinetics in agreement with the Michaelis-Menten term used 
for the complete reaction in the original full scale model~\cite{Hynne01}.
\begin{eqnarray}
v_{\mathrm{ADH,ACA}}^+ &=& k^{\mathrm{ADH}}_1\cdot[{\mathrm{ACA}}]\cdot[{\mathrm{NADH}}]\cdot[{\mathrm{ADH}}]  \\
v_{\mathrm{ADH,ACA}}^- &=& k^{\mathrm{ADH}}_2\cdot[{\mathrm{ADH}}_{\mathrm{ACA}}] +
k^{\mathrm{ADH}}_3\cdot[{\mathrm{ADH}}_{\mathrm{ACA}}]
\end{eqnarray}
The unbound form of the enzyme is computed from the total amount
$ADH_{\mathrm{total}}$ of the enzyme by
\begin{equation}
[{\mathrm{ADH}}]=ADH_{\mathrm{total}}-[{\mathrm{ADH}}_{\mathrm{ACA}}]
\label{equ:adh_constraint1}
\end{equation}
whereas the concentration of NAD$^+$ follows from
\begin{equation}
[{\mathrm{NAD}}^+]=N_{\mathrm{total}}-[{\mathrm{NADH]}}-[{\mathrm{ADH}}_{\mathrm{ACA}}]~.
\label{equ:nadh_constraint1}
\end{equation}
The ODEs for [NADH], [ACA] and [EtOH] have also been updated by the new
expressions for the fluxes.
We have fixed two of the four new parameters to $ADH_{\mathrm{total}}=10 \mu$M and
$k^{\mathrm{ADH}}_2=1000$ min$^{-1}$ and computed the remaining two self\sep consistently from 
the fluxes in the original full scale model~\cite{Hynne01}, 
for all values see Table~\ref{tab:par}.
This procedure preserved the full scale model's behavior as verified in 
Table~\ref{tab:var} and by comparison of the bifurcation diagrams
Fig.\ref{fig:bifglc} (a) and Fig.~5 in Ref.~\cite{Hynne01}.
Identical equations (\ref{equ:adh_aca_ode})-(\ref{equ:nadh_constraint1}) and the
same parameter values are used in both models.

Although the core model possesses the same topology as the full scale model, 
it uses simple mass action kinetics for all lumped steps of the
pathway~\cite{WH00}. One single regulatory interaction has been included, 
specifically the regulation of 6-phosphofructokinase (PFK; EC 2.7.1.11) via ATP.
This strong abstraction of the core model is sufficient to
reproduce glycolytic oscillations~\cite{WH00}.

\begin{table}
\begin{center}
\begin{tabular}{llrrr}
\hline
\scriptsize Variable in core & \scriptsize and full scale model & \scriptsize Core model & \scriptsize Full scale model & \scriptsize Reference \\
\hline
\scriptsize [Glc$_x$]$_0$&\scriptsize [Glc$_x$]$_0$& \scriptsize 18.5&   \scriptsize 18.501& \scriptsize 18.5\\[-3mm]
\scriptsize $S_1^{ex}$&\scriptsize [Glc$_x$]& \scriptsize 1.54982&  \scriptsize 1.55409& \scriptsize 1.55307\\[-3mm]
\scriptsize $S_1$ & \scriptsize [Glc]     &  \scriptsize 0.560999& \scriptsize 0.573935& \scriptsize 0.573074\\[-3mm]
\scriptsize $S_2$ & \scriptsize [DHAP]    &   \scriptsize 2.08167&  \scriptsize 2.97161& \scriptsize 2.95\\[-3mm]
\scriptsize $S_3$ & \scriptsize [BPG]     &\scriptsize 0.000266672&\scriptsize 0.00026973& \scriptsize 0.00027\\[-3mm]
\scriptsize $S_4$ & \scriptsize [ACA]     &   \scriptsize 1.48032&  \scriptsize 1.49748& \scriptsize 1.48153\\[-3mm]
\scriptsize $S_4^{ex}$&\scriptsize [ACA$_x$]& \scriptsize 1.28731&  \scriptsize 1.30226& \scriptsize 1.28836\\[-3mm]
\scriptsize $N_2$ & \scriptsize [NADH]    &  \scriptsize 0.330046& \scriptsize 0.329872& \scriptsize 0.33\\[-3mm]
\scriptsize $A_3$ & \scriptsize [ATP]     &   \scriptsize 2.08505&  \scriptsize 2.09875& \scriptsize 2.1\\
\hline
\scriptsize $j_{\mathrm{in}}$     & \scriptsize $j_{\mathrm{in}}$     &  \scriptsize 48.0& \scriptsize 47.99& \scriptsize 48.0\\[-3mm]
\scriptsize $j_{\mathrm{storage}}$& \scriptsize $j_{\mathrm{storage}}$&  \scriptsize 20.0& \scriptsize 19.90& \scriptsize 20.0\\[-3mm]
\scriptsize $j_{\mathrm{Glyc}}$   & \scriptsize $j_{\mathrm{Glyc}}$   &  \scriptsize 4.77&  \scriptsize 4.82&  \scriptsize 4.8\\[-3mm]
\scriptsize $j_{\mathrm{EtOH}}$   & \scriptsize $j_{\mathrm{EtOH}}$   & \scriptsize 46.46& \scriptsize 46.54& \scriptsize 46.4\\[-3mm]
\scriptsize $j_{\mathrm{outACA}}$ & \scriptsize $j_{\mathrm{outACA}}$ & \scriptsize 4.77&  \scriptsize 4.82&  \scriptsize 4.8\\
\hline
\end{tabular}
\end{center}
\caption{
Comparison of metabolite concentrations (top, all values are in mM and
accurate to six significant digits) and fluxes 
(bottom, rounded and in mM min$^{-1}$) 
for both models at the onset of oscillations.
Reference values are taken from a comprehensive modeling effort that
incorporated available experimental data (Table 6 in~\cite{Hynne01}) and only a subset of
the full scale model's variables are shown. All others match the Reference
equally well.}
\label{tab:var}
\end{table}

We calculated a new set of parameters (see Table \ref{tab:par}) for the core 
model such that the onset of oscillations at [Glc$_x$]$_0$=18.5 mM
is captured to the same accuracy (see Table \ref{tab:var}) 
as in the full scale model, which had particularly been optimized to 
describe the onset of oscillations.
To fit the core model, we inserted the known values of fluxes and concentrations
from the full scale model 
(Table 6 in Ref.~\cite{Hynne01}, here included as Reference in Table \ref{tab:var})
into the expressions of the core model and calculated
those parameters $k_i$ which are factors of mass action terms.
This strategy was partly used in Ref.~\cite{Hynne01}.
Parameters $k_i$ of flux terms that involve a yet undetermined 
saturation constant were expressed as functions 
(with inserted values of fluxes and concentrations at onset) 
of the respective saturation constant.
Then a steady state at the onset of oscillations (Hopf bifurcation) was 
determined for an initial choice of the unknown parameters and numerically 
continued along curves in the space of unknown parameters (see Methods below).
Note, this strategy is very efficient since each run explores two unknown
parameters simultaneously which are linked by the constraint of reproducing a 
Hopf bifurcation. 
We monitored the values of the resulting fluxes and fixed that
combination of remaining parameters with closest correspondence of the computed 
fluxes to the data from the full scale model 
(Tables 7,8 in Ref.~\cite{Hynne01}).
The computed concentrations at the onset of oscillations in the core model were
also in good agreement with those of the full scale model 
(see Table \ref{tab:var}).
The temporal period of the oscillations near onset was close to 0.65 min in the
full scale model and 0.25 min in the core model.

Fig.~\ref{fig:bifglc} shows that the new set of parameters also yields a 
supercritical onset of the oscillations in the core model 
in agreement with the full scale model and experiments~\cite{DSH99,Hynne01}.
This set of
parameters is therefore well suited for further studies of synchronisation
phenomena in a multicellular context~\cite{WPSSHW00}.
Note, that both models also share a second oscillatory domain at lower glucose
concentration. This has also been observed in the original full scale model
(Fig.~8 in Ref.~\cite{Hynne01}) but so far not in experiments.
This agreement of both models' predictions may stem from their identical 
topology but still is remarkable since it covers a parameter range outside of
that where the models had been fitted.

Altogether, both the full scale model and the core model show very similar
behavior and create the impression that quantitative details are largely
irrelevant when the model is analysed in the same context that had already been
used to determine the remaining unknown parameters. 
Hence ``predictions'' from a simplified model are reliable for conditions
similar to those considered in the model's construction, {\it e.g.}
both models predict a second oscillatory domain for low glucose supply.
However, we want to test the reliability of the core model under conditions that
may be very different from those included during parameter optimization.
Such reliability is essential for the rational improvement of biotechnological
processes beyond the conditions applied so far.

\subsection{Perturbation of the redox-balance}

Both the full scale model and the core model have been augmented with the same
set of equations that account for the biotransformation of a xenobiotic ketone.
In the considered case the ketone ethyl acetoacetate is reduced by alcohol 
dehydrogenase (ADH; EC 1.1.1.1) to enantiomerically pure ethyl 
$L$-3-hydroxybutyrate (carbinol).
Although this reaction has been studied extensively, little is known about how the 
intracellular enzymes interact~\cite{BGSDN02,GFLS00,DM98,N96,MSO94,KKM91}.
Currently, two 
enzymes, $D$-directing $\beta$-ketoacyl reductase (KAR; EC 1.1.1.100) and $L$-directing  
ADH are known to be involved in the asymmetric reduction of 
the substrate~\cite{B02,SSJPH98}. Yet, KAR activity is negligible with a large excess of 
sugar present. Hence, the overall stereo-selectivity of the bioconversion largely depends on 
cell physiology~\cite{B02}. 

In analogy to the reduction of acetaldehyde (ACA) in
reaction~(\ref{equ:adh_aca}),
we now consider the reduction of ketone (K) to carbinol (C) 
by the same simple reaction scheme:
\begin{equation}
\begin{array}{lc}
\mbox{K + NADH + ADH} &
\begin{array}{c}
k^{\mathrm{ADH}}_4 \\[-3mm]
\rightleftharpoons \\[-2mm]
k^{\mathrm{ADH}}_5
\end{array}\\[5mm]
\mbox{ADH$_{\mathrm{K}}$} \qquad 
\begin{array}[b]{c}
k^{\mathrm{ADH}}_6 \\[-4mm]
\to
\end{array}
\qquad \mbox{ADH + NAD$^+$ + C}
\end{array}
\label{equ:adh_k}
\end{equation}
with rate constants $k^{\mathrm{ADH}}_4, k^{\mathrm{ADH}}_5$ and $k^{\mathrm{ADH}}_6$.
If not stated differently then the new parameters are chosen equal to the
corresponding values of the acetaldehyde reduction, 
{\it i.e.} $k^{\mathrm{ADH}}_4=k^{\mathrm{ADH}}_1, 
k^{\mathrm{ADH}}_5=k^{\mathrm{ADH}}_2$ and 
$k^{\mathrm{ADH}}_6=k^{\mathrm{ADH}}_3$.
The corresponding ODE with gain ($v_{\mathrm{ADH,K}}^+$) and loss
($v_{\mathrm{ADH,K}}^-$) fluxes 
reads
\begin{eqnarray}
\frac{\rm d [{\mathrm{ADH}}_{\mathrm{K}}]}{\rm dt} &=& v_{\mathrm{ADH,K}}^+ -
v_{\mathrm{ADH,K}}^- \\
v_{\mathrm{ADH,K}}^+ &=& k^{\mathrm{ADH}}_4\cdot[{\mathrm{K}}]\cdot[{\mathrm{NADH}}]\cdot[{\mathrm{ADH}}] \\
v_{\mathrm{ADH,K}}^- &=& k^{\mathrm{ADH}}_5\cdot[{\mathrm{ADH}}_{\mathrm{K}}] +
k^{\mathrm{ADH}}_6\cdot[{\mathrm{ADH}}_{\mathrm{K}}]~.
\end{eqnarray}
The additional variable [ADH$_{\mathrm{K}}$] also alters the constraints and 
yields via the consumption of [NADH] a strong feedback on the upstream steps 
of glycolysis.
\begin{eqnarray}
[{\mathrm{ADH}}] &=& ADH_{\mathrm{total}}-[{\mathrm{ADH}}_{\mathrm{ACA}}]-[{\mathrm{ADH}}_{\mathrm{K}}] \\
\label{equ:adh_constraint2}
[{\mathrm{NAD}}^+] &=& N_{\mathrm{total}}-[{\mathrm{NADH}}]-[{\mathrm{ADH}}_{\mathrm{ACA}}]-[{\mathrm{ADH}}_{\mathrm{K}}] 
\label{equ:nadh_constraint2}
\end{eqnarray}
The membrane transporters for ketone 
$v_{\mathrm{K}} = \kappa \cdot ([{\mathrm{K}}_x]-[{\mathrm{K}}])$ 
and carbinol were chosen linear in order 
not to restrict the accessible range of intracellular perturbations.
The permeability for ketone $\kappa=24.7\ \mbox{min}^{-1}$ 
was set equal to the value for acetaldehyde \cite{Hynne01}.

The above perturbation introduces a competition between acetaldehyde and ketone 
for unbound enzyme ADH and the cofactor NADH.
In Sect.~\ref{results}, the behavior of both models is tested for consequences 
of the perturbation and predictions on oscillatory or steady behavior are 
compared.

\subsection{Methods}

The investigated metabolic network may be considered as a nonlinear dynamical
system that evolves its own state, {\it i.e.} the set of values for the
metabolites' concentrations, in a self\sep organizing manner.
Various computational methods have been developed and proven to be essential 
for the analysis of complex cellular networks \cite{TCN01,FMWT02}, {\it e.g.}
bifurcation analysis \cite{S94,KG95} and direct numerical simulation 
\cite{Mendes97}.

Since we focus on steady and oscillatory states
it is most convenient to combine the computation of the system's state 
along ``branches'' in order to efficiently scan the parameter space.
This strategy is called ``continuation'' and starts from a single known solution
and proceeds in subsequent steps to compute similar solutions for neighboring 
values of the parameters, {\it e.g.} [Glc$_x$]$_0$ (mixed flow glucose
concentration at inlet), [K$_x$]
(extracellular ketone concentration).
We used the software package AUTO for such tasks \cite{AUTO}.
Therein the solutions are represented 
as the root of an extended system of algebraic equations derived by
discretizing the ODEs. Any new solution (the root) is then computed 
iteratively by the Newton algorithm.

The software also monitors the linear stability of the solutions and 
thereby detects critical values of the parameters
where the solutions change drastically, so called ``bifurcations''.
The onset of oscillations, a Hopf bifurcation, is of special interest here
hence the employed strategy amounts to a bifurcation analysis.
Bifurcations may themselves be continued in two control parameters 
simultaneously which was used to fit the parameter set of the core model to 
flux and concentration data of the full scale model at the Hopf bifurcation.
We have supplemented the AUTO package by subroutines for oscillatory states 
that monitor and export the time averages of fluxes and the relative phase 
shifts between periodic time courses of concentrations.


\begin{figure}
\centerline{\includegraphics[width=.99\textwidth]{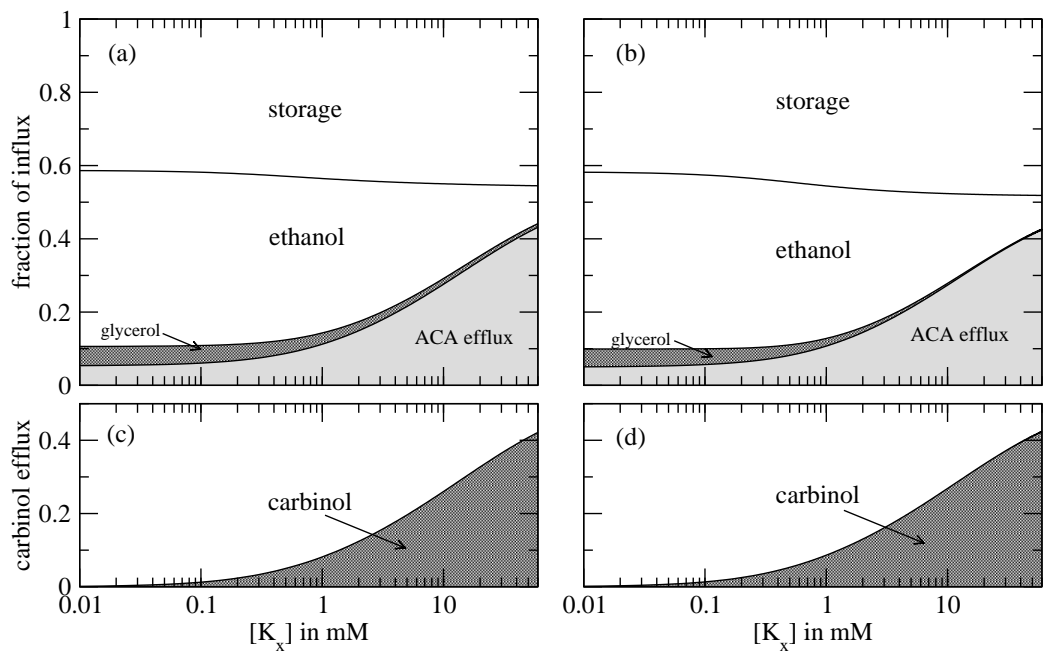}}
\caption{\label{fig:ketx_flux} 
Flux diagrams (log\sep linear scales) at fixed glucose concentration 
[Glc$_x$]$_0$=30mM for (a) the full scale model and (b) the core model.
All fluxes have been measured in equivalents of C$_3$ entities and have been 
normalized by the influx (96.2 mM$_{\mathrm{C}_3}$ min$^{-1}$).
Individual pathways (storage, ethanol production, glycerol production, 
acetaldehyde (ACA) efflux as annotated) consume a portion of the influx given 
by the distance between the curves above and below the respective area.
(c),(d) The biotransformation flux for both models.
}
\vspace*{2mm}
\end{figure}

\section{Results}\label{results}

\subsection{Comparison of model predictions}

The fluxes within the metabolic network of glycolysis are redirected when 
the extracellular ketone concentration is increased.
Both models show almost identical behavior, see Fig.~\ref{fig:ketx_flux}.
The effluxes (storage, ethanol, glycerol, acetaldehyde) are represented by the
intervals between the curves above and below the corresponding area and all
effluxes add up to the total influx in units of C$_3$.
The observed increase in the efflux of acetaldehyde
corresponds to the additional consumption of cofactor NADH via ketone reduction.
The latter is fostered by increasing ketone concentration, and the stoichiometric
constraint on NADH recycling (light shaded area equals sum of dark shaded areas) 
demands an equal flux of NADH regeneration.
Glycerol secretion is discriminated by the lack of NADH, however, in the full
scale model a small flux to glycerol remains even for large ketone concentration
whereas this flux is strongly suppressed in the core model.
We observed the same behavior independently of the value of [Glc$_x$]$_0$.
Fluxes have been averaged over time for the oscillatory states. 
Note, the time courses of substrates and cofactors in oscillatory states are 
generally out of phase which influences the average fluxes, see
Subsec.~\ref{subsec:phaseshift}.

\begin{figure}
\vspace*{2mm}
\centerline{\includegraphics[width=.99\textwidth]{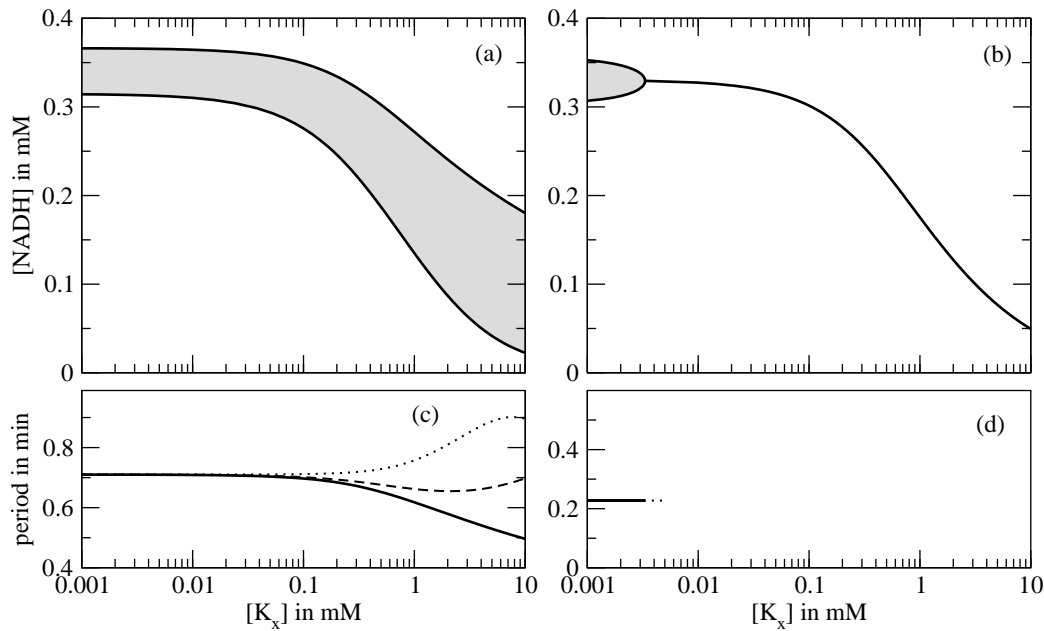}}
\caption{\label{fig:ketx_ampl} 
Bifurcation diagrams for (a),(c) the full scale model and (b),(d) the core
model at fixed [Glc$_x$]$_0$=30mM. 
Oscillations of the NADH concentration (gray area in (a),(b)) occur 
between a maximum and a minimum value (solid curves) that decrease for 
stronger perturbation by ketone.
(c),(d) Temporal period of the oscillations for $k^{\mathrm{ADH}}_6 = 5400$
min$^{-1}$ (solid curve) that decreases (frequency increases) with [K$_x$] for
the full scale model in (c).
Additionally, data for $k^{\mathrm{ADH}}_6 = 4800$ min$^{-1}$ (dashed) and 
$k^{\mathrm{ADH}}_6 = 3800$ min$^{-1}$ (dotted) indicate that periods may also
remain unaffected or increase with [K$_x$] depending on the carbinol release 
rate.
}
\end{figure}

The dependence of the oscillations on the strength of the
ketone reduction is shown in Fig.~\ref{fig:ketx_ampl}.
Both models behave differently with sustained oscillations predicted by
the full scale model over the whole range of perturbations whereas the core
model does favor steady behavior.
However, the temporally averaged concentration of NADH shows the same 
depression upon ketone increase, reflecting the additional NADH consumption.

We repeated the analysis for several values of the carbinol release rate 
$k^{\mathrm{ADH}}_6$ from the ADH complex, see Eq.~(\ref{equ:adh_k}).
Different release rates are expected for structurally dissimilar 
ketones due to varying interactions between the enzyme and structures 
besides the common keto-group.
The corresponding variability in the affinity $k^{\mathrm{ADH}}_4$ 
and in $k^{\mathrm{ADH}}_5$ are not considered as such changes
can be compensated by rescaling of [K$_x$] and do not yield new results.

Fig.~\ref{fig:ketx_ampl} (c) shows how the oscillation period is affected by
ketone for three choices of the release rate $k^{\mathrm{ADH}}_6$.
The oscillations speed up (period decreases) for 
$k^{\mathrm{ADH}}_6 > k^{\mathrm{ADH}}_{6cr} = 4800$ min$^{-1}$ 
including the case of equal release rates 
$k^{\mathrm{ADH}}_3 = k^{\mathrm{ADH}}_6$ and the oscillations slow down
(period increases) if the release rate is chosen slower than 
$k^{\mathrm{ADH}}_{6cr}$.
The critical rate $k^{\mathrm{ADH}}_{6cr}$ arises from the competition of
additional NAD$^+$ recycling flux due to additional substrate (ketone) versus
diminished NAD$^+$ recycling via ACA reduction when the enzyme is trapped longer
in the ADH$_\mathrm{K}$ complex and therefore less active.
The former effect accelerates the oscillations whereas the latter slows them 
down.
Sustained oscillations have been observed over long time scales \cite{DSH99}
which enable accurate measurements of the oscillation period.
Such measurements can be used to infer rate constants of different ketones
relative to the rates of ACA reduction and to distinguish different ketones.

\begin{figure}
\vspace*{2mm}
\centerline{\includegraphics[width=.99\textwidth]{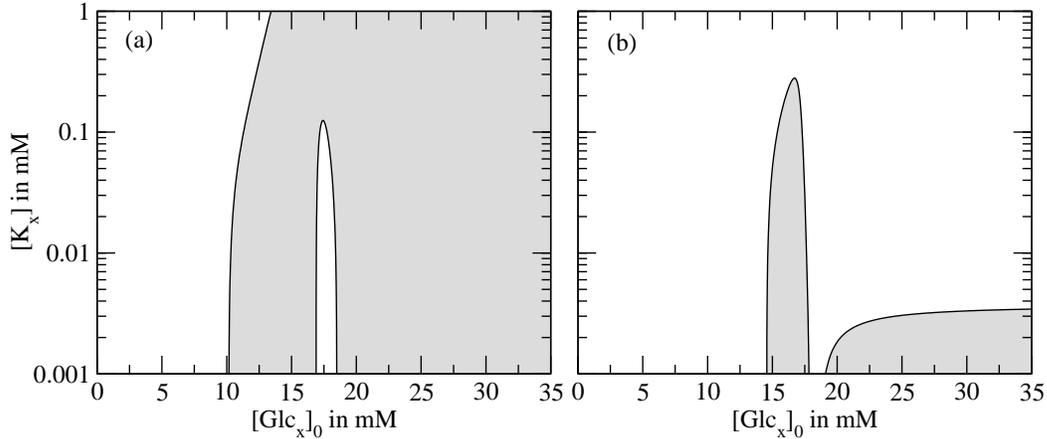}}
\caption{\label{fig:ketx} 
Phase diagram of stable stationary (white) and oscillatory (gray) states in the
[Glc$_x$]$_0$ - [K$_x$] parameter space (linear\sep logarithmic scales)
for (a) the full scale model and (b) the core model.
}
\end{figure}

Fig.~\ref{fig:ketx_ampl} (d) confirms that the core model does not possess the
dynamical features discussed above.
In order to characterize the conflicting predictions of both models, we have
computed the parameter values of the onset of oscillations in a two\sep parameter
plane, see Fig.~\ref{fig:ketx}. We observe a qualitative difference in terms of
the relations between the three Hopf bifurcations that are present in the 
unperturbed system.
The full scale model possesses extended oscillatory domains (gray shaded). 
When [K$_x$] is increased, two Hopf bifurcations at {\em large} 
[Glc$_x$]$_0$ are connected and the enclosed steady behavior vanishes.
The core model favors stationary states: two
Hopf bifurcations at {\em small} [Glc$_x$]$_0$ are connected
and the enclosed oscillations vanish when [K$_x$] is increased.
Hence, the different levels of abstraction mainly determine the dynamical aspects
of the models' behavior. 
On the other hand the averaged or stationary quantities are much less affected
by the considered abstraction.

We further examined the case of low membrane permeability for ketone or 
equivalently low external concentration of ketone 
(to maintain viability) which both yield 
small intracellular concentrations of ketone.
At a fixed ketone concentration ([K$_x$] $<$ 1 mM), we varied [Glc$_x$]$_0$ 
and monitored the fluxes.
A maximum of carbinol production is found at an intermediate value of 
[Glc$_x$]$_0$ which depends on the fixed value [K$_x$] (not shown).
Both models predict this maximum which can be interpreted as follows.
The lack of NADH recycling is limiting the carbinol flux for small 
[Glc$_x$]$_0$ whereas at large [Glc$_x$]$_0$ almost all ADH binds to acetaldehyde
which is then much more abundant than the competing ketone.
Hence at intermediate value an optimum carbinol flux develops where both 
limiting influences are weak. In biotechnological applications it is often 
desired to adjust the glucose supply accordingly.

\subsection{Phase shift in oscillatory states}\label{subsec:phaseshift}

Finally, we have closer analysed possible differences between temporally
averaged oscillatory fluxes and the coexisting steady state fluxes, both in the 
full scale model. Note, the steady states are unstable when oscillations occur 
supercritically but unstable states can nevertheless be computed by the
continuation algorithm as this method does not rely on temporal integration.
The unstable steady state represents a solution that equally respects
stoichiometric and regulatory constraints as the stable oscillatory state does
and a minor change in parameters can render the steady state stable. 
Hence the coexisting unstable steady state reproduces the behavior of an 
alternative operational mode.

\begin{figure}
\vspace*{2mm}
\centerline{\includegraphics[width=.99\textwidth]{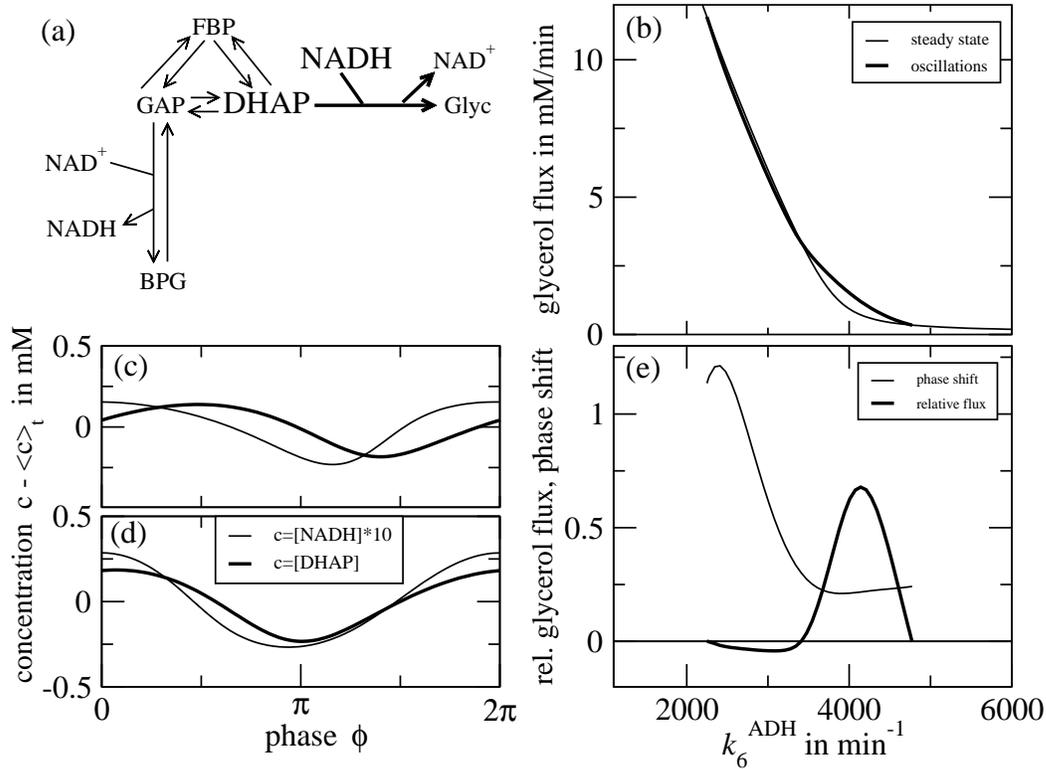}}
\caption{\label{fig:phase} 
(a) Small portion of the reaction scheme of glycolysis corresponding to the
shaded box in Fig.~\ref{fig:sketch}(b). The substrate DHAP is
reduced by NADH to finally yield glycerol.
(b) Temporally averaged glycerol flux $v_{\mathrm{osc}}$ in oscillatory 
state (thick curve between two Hopf bifurcations at 
$k^{\mathrm{ADH}}_6 \approx$ 2250 min$^{-1}$ and 4750 min$^{-1}$) 
is found to be larger or smaller, respectively, than the corresponding 
steady state flux $v_{\mathrm{steady}}$ (thin curve).
(c) Oscillatory time courses of [NADH] (thin) and [DHAP] (thick curve) 
are out of phase by $\Delta \phi=1.2$ at $k^{\mathrm{ADH}}_6$=2300 min$^{-1}$.
Average concentrations $\langle c \rangle_t$ have been subtracted and [NADH] has been scaled
by a factor of 10 in (b,c).
(d) Both concentrations oscillate almost in phase ($\Delta \phi=0.2$) at larger
$k^{\mathrm{ADH}}_6$, here 4600 min$^{-1}$.
(e) The relative difference of fluxes
$(v_{\mathrm{osc}}-v_{\mathrm{steady}})/v_{\mathrm{steady}}$ (thick) 
is partly due to varying phase shift $\Delta \phi$ (thin curve) between 
the time courses of substrate DHAP and cofactor NADH. 
Parameters are [Glc$_x$]$_0$=15mM and [K$_x$]=10mM.
}
\end{figure}

As an example, we compare the fluxes towards glycerol production for a range of
values of the carbinol release rate $k^{\mathrm{ADH}}_6$ from the ADH complex, 
see the remarks above and Eq.~(\ref{equ:adh_k}).
Fig.~\ref{fig:phase} (a) shows a small portion of the reaction scheme of
glycolysis where dihydroxyacetone phosphate (DHAP) and NADH are reaction 
partners. The concentrations of fructose 1,6-bisphosphate (FBP), 
glyceraldehyde 3-phosphate (GAP) and DHAP perform synchronized oscillations
whereas the cofactors NADH and NAD$^+$ always possess anti-cyclic oscillations.
Therefore, this branch point of glycolysis (Fig.~\ref{fig:phase} (a) and
shaded area in Fig.~\ref{fig:sketch} (b)) may act
as a switch if the oscillation of either NADH or NAD$^+$ can be synchronized
with the substrate pool.

The computations are performed in the full scale model and all parameters of the
glycerol path are kept fixed.
Fig.~\ref{fig:phase} (b) shows the averaged oscillatory glycerol flux that is
increased by more than 50\% at large rate $k^{\mathrm{ADH}}_6$ 
with respect to the steady state at the same rate.
The time courses of substrate DHAP and cofactor NADH are depicted in 
Fig.~\ref{fig:phase} (c,d) for two different values of $k^{\mathrm{ADH}}_6$.
The temporally averaged concentrations $\langle c \rangle_t$ are almost
identical with the steady state value at the same $k^{\mathrm{ADH}}_6$.
We define a phase $\phi(c)\equiv 2\pi (t(c_{max})+t(c_{min}))/2T$
to characterize a time course with period $T$ and time of maximum
($t(c_{max})$) and minimum concentration ($t(c_{min})$).
The large increase in averaged oscillatory flux versus the steady state value
stems from the oscillations of DHAP and NADH that are almost in phase at 
large $k^{\mathrm{ADH}}_6$ (Fig.~\ref{fig:phase} (d,e)).

Slower release of NAD$^+$ (smaller $k^{\mathrm{ADH}}_6$) increases
the phase shift ($\Delta \phi \equiv \phi([$DHAP$])-\phi([$NADH])) 
to the trailing DHAP oscillation (Fig.~\ref{fig:phase} (c)) and 
decreases the averaged oscillatory flux relative to the steady state flux
(Fig.~\ref{fig:phase} (e)).
In addition to the phase shift $\Delta \phi$, the averaged oscillatory flux is
also affected by changes (due to varied $k^{\mathrm{ADH}}_6$) 
in amplitude and functional form of the oscillations.
Moreover, the absolute values of both steady and averaged oscillatory flux 
vary strongly (Fig.~\ref{fig:phase} (b)) as the temporally
averaged metabolite concentrations change with $k^{\mathrm{ADH}}_6$.
The biotransformation flux towards carbinol is affected by changes of similar 
absolute size which, however, amount to slight relative changes only.

In the left half of the oscillatory interval, the averaged oscillatory flux 
to glycerol is even smaller than in steady state which is only possible for 
a large phase shift, as observed in Fig.~\ref{fig:phase} (c).
The effect of the phase shift becomes dominant for large oscillation 
amplitudes because it clearly matters, if the
reaction partners frequently collide when they appear at the same time or if 
they only rarely react when one of them is always too late.
This situation is analogous to a problem in electrical engineering
where the electrical power at a device in a sinusoidal alternating current (AC) 
depends on the amplitudes of voltage and current and on the phase shift between
their maxima.

\section{Conclusions and perspective}\label{conclusions}

We have compared predictions of two theoretical models that have previously 
been used to describe glycolytic oscillations in yeast. Both models possess the
same topology but incorporate different levels of kinetic details. 
Irrespective of these differences, both models predicted identical dependencies
of temporally averaged fluxes and concentrations on the strength of the
perturbation, {\it i.e.} the extracellular concentration of a xenobiotic
ketone. This observation provides confidence in theoretical modeling given the
topology of the network is known and completely incorporated. Note, the
unknown parameters of the models were fitted to data of the unperturbed system
hence the models predicted the response to the perturbation. 
The chosen perturbation affected the entire network via multiple feedback loops
of cofactors. We suggest the exposure to xenobiotics as a feasible experimental 
probe of the network's global response. In the future we will extend this
analysis to metabolic responses in other eucaryotes and to a range of 
pharmacologically active compounds.

The kinetic details of both models affected the temporal organisation of the
responses. Oscillations ceased to exist in the abstract core model while the
full scale model supported oscillations for a large range of perturbations.
Oscillatory behavior has recently been suggested by Hauser {\it et al.} to 
protect enzymes and possibly cells against toxic 
substances~\cite{HKLO01,OHK03}. 
Oscillations also play an important role for calcium
signal transduction~\cite{SignalTrans} where they appear in
bursts~\cite{CaBurst,FoldHopf}.
Bursting, the temporally repeated sequence of quiescent and oscillating
episodes, has recently been found in a theoretical study of 
stochastically driven glycolysis~\cite{KR03}.
Oscillations have also been suggested to improve the thermodynamic
efficiency of glycolysis with respect to steady operation~\cite{RR80}.
Furthermore, kinetic details are important for (non periodic) 
short term responses to glucose pulses~\cite{Rizzi97}.

A closer inspection of the flux distribution in oscillatory states revealed 
that temporally averaged fluxes may differ from the values in the 
coexisting (unstable) steady state, 
albeit for each metabolite the (unstable) steady state concentration equals
the temporally averaged value in the oscillatory state.
This effect results from the relative phase shift between time courses of
substrates and cofactors particularly at branch points of the network.
The time average of any oscillatory flux is generally restricted 
to a convex combination of elementary flux modes \cite{SDF99}. 
However, allosterically regulated enzymes impose additional 
restrictions on the fluxes at given metabolite concentrations which can
result in suboptimal convex combinations particularly in steady states.
Oscillatory states possess additional degrees of freedom, {\it i.e.} phase
shifts, that are adjusted by many even remote reactions.
Evolution of metabolic networks may therefore have acted faster on those
networks capable of oscillatory behavior enabling them to explore 
more possibilities and to become superior.
This conjecture is in line with the observed oscillatory behavior in one of the
most important and inherited metabolic pathways, {\it i.e.} glycolysis, and it
supports the view that oscillatory behavior may be more abundant than 
presently recognised \cite{HKLO01,OHK03}.

A corresponding strategy for metabolic engineering may target the phase shifts
in oscillatory states as has been suggested by Ross and coworkers \cite{LR90}.
They forced simple reaction schemes with oscillating substrate concentrations
to increase desired fluxes.
Here we found that the same effect is also important for large metabolic 
networks where phase shifts are adjusted in a self-organizing fashion.
The artificial addition of appropriate buffers or bypasses (as in the present
example) in order to control phase shifts are promising candidates for 
metabolic engineering.

In terms of modeling strategies, one should treat those approaches with 
caution that focus on steady states, {\it e.g.} canonical modeling
\cite{V00}, since the computed flux distribution in an unstable
steady state does not necessarily approximate the actual oscillatory 
behavior around that steady state.
Continuation algorithms as applied in the present paper appear to be 
particularly suited to address the (dis)advantages of oscillatory versus
(hypothetical) stationary behavior.

To summarize, we have shown that
static features of the metabolic network of glycolysis 
are robust to crude abstractions of models but more care
has to be taken when dynamic features are considered.
Many open questions remain and call for further research work. 
E.g. the possibility to conduct ketone-to-carbinol biotransformations under
aerobic conditions could increase the NADH recycling and hence the carbinol 
production which, \textit{per se}, is an important biotechnological challenge. 
Further modeling will help to solve the related problems by
simulating the effect of suppressing or enhancing parts of the complex 
metabolic network.

We are indebted to Markus B\"ar, Sune Dan{\o}, 
Jean-Marie Fran\c{c}ois, Ursula Kummer, Karl-Heinz van
P{\'e}e, Preben Graae S{\o}rensen and Jana Wolf for fruitful discussions. 
MB wishes to thank Gerhard J\"org for practical assistance
and the Graduate Research Center "Heterocycles" 
as well as the German National Merit Foundation for financial support.

\end{document}